\documentclass[12pt]{article}
\usepackage{amsmath, amssymb, graphicx}
\usepackage{geometry}
\usepackage{hyperref}
\usepackage{setspace}
\usepackage{caption}
\usepackage{algorithm}
\usepackage{algorithmic}
\usepackage{placeins}
\usepackage{comment}
\usepackage[authoryear,round]{natbib}
\setcitestyle{authoryear,round,aysep={},sep={;}}
\usepackage[english]{babel}
\usepackage[nottoc]{tocbibind}
\usepackage{subcaption}
\usepackage{ragged2e}
\title{Bibliography management: BibTeX}
\title{Bayesian inference for ordinary differential equations models with heteroscedastic measurement error}

\author{
Selva Salimi$^{1,2}$\thanks{Corresponding author: selva.salimi@hdr.qut.edu.au},
David J. Warne$^{1,2,3}$,
Christopher Drovandi$^{1,2,3}$ \\
\\
}
\date{
$^{1}$School of Mathematical Sciences, Faculty of Science, 
Queensland University of Technology, Brisbane, Australia. \\ 
$^{2}$Centre for Data Science, 
Queensland University of Technology, Brisbane, Australia. \\
$^{3}$ARC Centre of Excellence for the Mathematical Analysis of Cellular Systems, 
Queensland University of Technology, Brisbane, Australia.
}

\begin{document}

\maketitle

\begin{abstract}
Ordinary differential equation (ODE) models are widely used to describe systems in many areas of science, including biology, ecology, and epidemiology. To ensure these models provide accurate and interpretable representations of real-world dynamics, it is often necessary to infer parameters from data, which involves specifying the form of the ODE system as well as a statistical model describing the observational process.  A popular and convenient choice for the error model is a Gaussian distribution with constant variance (homoscedastic).  However, the choice may not be realistic in many systems, since the variance of the observational error may vary over time or have some dependence on the system state (heteroscedastic), reflecting changes in measurement conditions, environmental fluctuations, or intrinsic system variability.  Misspecification of the error model can lead to substantial inaccuracies of the posterior estimates of the ODE model parameters and predictions. More elaborate parametric error models could be specified, but this would increase computational cost because additional parameters would need to be estimated within the MCMC procedure and may still be misspecified. In this work we propose a two-step semi-parametric framework for Bayesian parameter estimation of ODE model parameters when there exists heteroscedasticity in the error process. The first step applies a heteroscedastic Gaussian process to estimate the time-dependent error, and the second step performs Bayesian inference for the ODE model parameters using the estimated time-dependent error estimated from step one in the likelihood function. Through a simulation study and two real-world applications, we demonstrate that the proposed approach yields more reliable posterior inference and predictive uncertainty compared to the standard homoscedastic models. Although our focus is on heteroscedasticity, the framework could be applied to handle more complex error processes.

\end{abstract}

\noindent\textbf{Keywords:} Time-varying error, Dynamical system modeling, Uncertainty quantification, Semi-parametric inference, Parameter estimation, Markov chain Monte Carlo

\section{Introduction}
Ordinary differential equation (ODE) models are widely applied across scientific disciplines, providing mechanistic descriptions of dynamical systems. Examples include the spread of infectious diseases in populations and cellular dynamics in biological systems \citep{Alahmadi2020,Browning2020,Maclaren2017}, interactions in ecological systems \citep{Warne_2025,https://doi.org/10.1111/1365-2664.14039}, supply-demand dynamics in economic systems \citep{XIA2026117428}, physics\citep{Potter2023} and classical physical processes such as Newton's law of cooling \citep{DAVIDZON20125397}. In each case, the dynamics of the system is determined by its current state together with a set of parameters that describe key features of the process, including rates, carrying capacities, and initial conditions. These parameters are rarely known a priori, and are often not directly measurable, necessitating their estimation from observational data. Reliable parameter estimation is therefore essential for accurate prediction and  understanding underlying processes.

A general ODE model can be written as
\begin{equation}
\frac{\textrm{d}y}{\textrm{d}t} = h(y, t; \theta) \quad \textrm{for } t>t_0 , \qquad y(t_0) = y_0 ,
\label{eq:ODE}
\end{equation}
where  $y(t) \in \mathbb{R}^n$ denotes the system state at time $t \geq t_0$, $\theta \in \mathbb{R}^m$ is a vector of unknown parameters, $h: \mathbb{R}^n \times \mathbb{R}^{+} \times \mathbb{R}^m \to \mathbb{R}^n$ is a function that defines dynamics of the system, that is the functional dependence of the rate of change on the state, $y$, time, $t$, and parameters, $\theta$, and $y_0$ is the initial condition. The observed data are denoted by $\{\tilde{y}(t_i)\}_{i=1}^N$, where $N$ is the number of observations, $t_i$ is the time of the $i$-th observation, and $\tilde{y}(t_i) \in \mathbb{R}^d$. Observations are typically linked to the latent state via
\begin{equation}
\tilde{y}(t_i) = g(y(t_i)) + \epsilon(t_i),
\label{eq:obs_model}
\end{equation}
where $g :\mathbb{R}^n \to \mathbb{R}^d$ maps the ODE solution to the same space on which data are observed. Here, \(d \le n\) indicates that the observations may represent fewer dimensions than the system state, and in some cases, none of the state variables may be directly observed. Further, $\epsilon(t_i) \in \mathbb{R}^d$ represents additive measurement error.

A standard assumption is that the errors are independent and identically distributed (i.i.d.) according to a Gaussian distribution \citep{Alahmadi2020,Murphy2022},
\[
\epsilon(t_i) \overset{\text{iid}}{\sim} \mathcal{N}(0,\sigma^2 \mathbb{I}),
\]
where $\mathbb{I}$ denotes the $d \times d$ identity matrix. In much of the existing literature, this i.i.d.\ Gaussian assumption is often adopted for analytical and computational convenience. However, real-world measurements may depart from satisfying such simplicity; errors can be heteroscedastic \citep{Warne_2025}, non-Gaussian \citep{NANSHAN2022107483}, or autocorrelated \citep{Lambert2023}, due to environmental variability or finite resolution of measuring instruments. Ignoring these complexities can lead to biased parameter estimates and inaccurate uncertainty quantification \citep{RePEc:bla:jorssc:v:71:y:2022:i:5:p:1978-1995,Warne2025.12.21.695338}. 

Recent work has begun to relax these simplifying assumptions. Non-Gaussian observation models have been explored using generalized linear frameworks \citep{NANSHAN2022107483}, while temporal dependence has been incorporated via parametric autocorrelation structures in Bayesian ODE models \citep{Lambert2023}. Although these approaches extend classical formulations, they typically rely on fully specified parametric error models and impose structural assumptions on the error process. 
Heteroscedasticity has also emerged as an important practical feature. For example, Warne et al. \citep{Warne_2025} consider observation error arising from the monitoring process, highlighting the importance of modeling assumptions and computational complexity in uncertainty quantification. These findings suggest that heteroscedastic error is not a minor complication but a structurally important feature of  time-series, motivating a dedicated focus on heteroscedasticity in our framework.

Heteroscedasticity can arise in multiple contexts. For example, model misspecification such as omitting relevant variables, assuming an incorrect functional form, or mischaracterizing the deterministic or stochastic nature of the system, can produce residuals whose variance depends on the state of the system \citep{hetarticle}. Similarly, measurement errors can introduce heteroscedasticity,  unusually small or large observations can influence the residuals more strongly than typical observations \citep{hetarticle}. Our work develops a framework to address the impact of incorrect measurement error assumptions on parameter estimation and predictive uncertainty in ODE models.

\begin{figure}[ht!]
\centering
\includegraphics[width=1\textwidth]{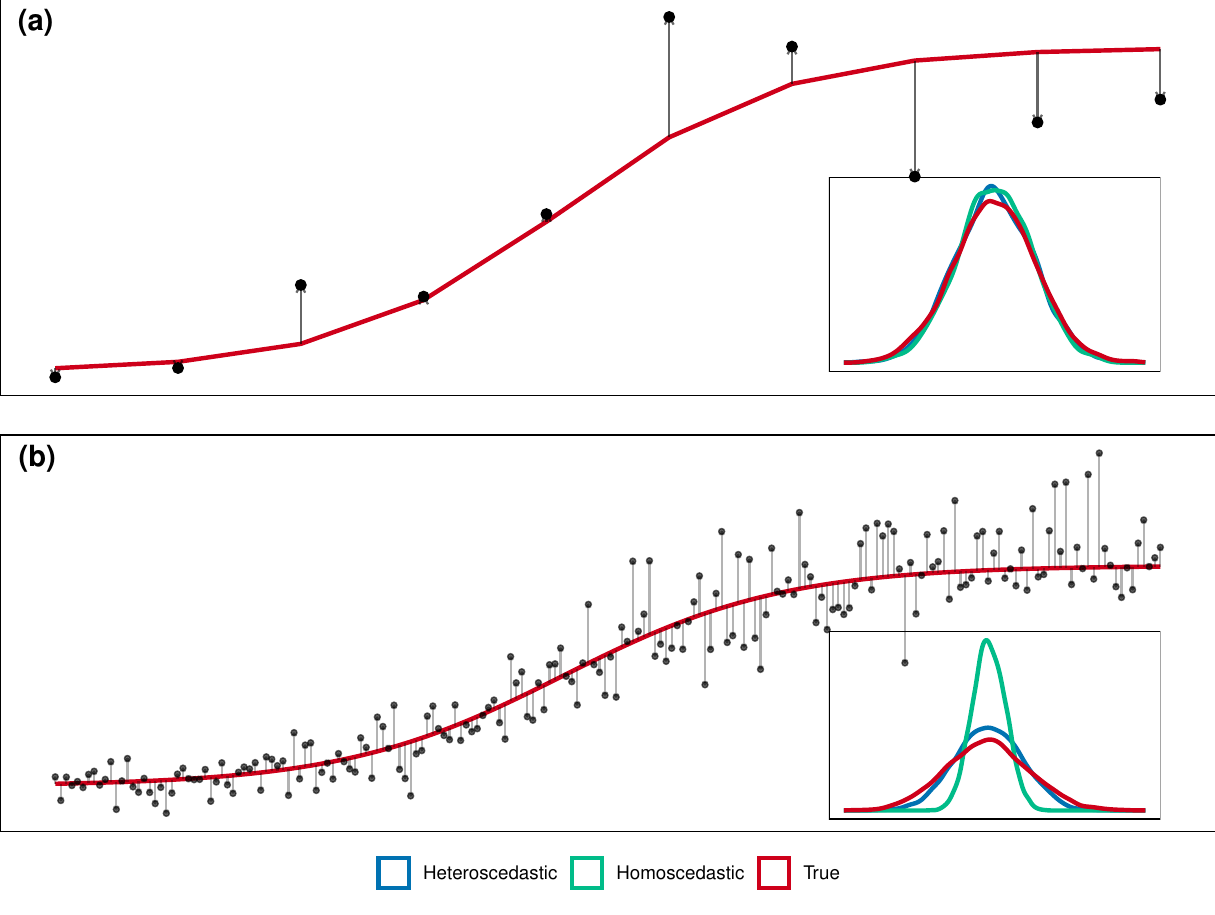}
\captionsetup{
    width=0.9\textwidth,
    font=footnotesize,  
    labelfont=bf,      
    justification=justified,
    skip=6pt         
}
\caption{(a) Sparse observations: With few data points, the posterior distributions under homoscedastic and heteroscedastic error assumptions are similar, as limited data provides little information to detect time-varying variance. (b) Dense observations: With more data points, heteroscedastic models capture time-varying error, revealing that homoscedastic assumptions can lead to overconfident or underconfident parameter estimates, particularly in regions of high or low measurement variability. Heteroscedastic models therefore provide more accurate uncertainty quantification across the input space}
\label{fig:motivation}
\end{figure}

Figure~\ref{fig:motivation} illustrates the practical implications of heteroscedasticity.
Figure~\ref{fig:motivation}a shows the scenario with relatively few observations. Posterior distributions obtained under the true model, an incorrect homoscedastic model that assumes constant variance, and our proposed approach (introduced in Section~\ref{our_method}) are largely similar. In this sparse data setting, limited information masks the underlying variability in the error process, making the simplifying constant-variance assumption appear adequate.
By contrast, in Figure~\ref{fig:motivation}b, increased observation density exposes the limitations of the homoscedastic assumption. The posterior obtained under constant variance is excessively tight, reflecting overconfident parameter estimates. The true model captures the increasing variability over time, while our approach produces posterior distributions that closely align with the true model and more accurately reflect the variability in the data.

Motivated by these considerations, in this work, we develop a semi-parametric Bayesian inference framework for ODE models that accommodates heteroscedastic measurement error. Our two-step approach leverages heteroscedastic Gaussian processes (HetGP)  \citep{article_HetGP} to learn the structure of the error process directly from data, separating the mechanistic dynamics from the stochastic variability. Importantly, this approach removes the need to assume a specific error model, since the error structure is automatically learned from the data. The estimated error process is then incorporated into the Bayesian inference procedure via the likelihood function, enabling parameter estimation. More accurately capturing the true error process in a semi-automated manner allows for more accurate Bayesian inference of the ODE model parameters, which are typically of main interest in scientific applications. Through applications to both synthetic and real datasets, we demonstrate that this framework yields more reliable inference across a range of scenarios compared with standard i.i.d.\ Gaussian assumptions. 

This separation between the mechanistic model and the stochastic component is novel and practically valuable. It allows for flexible adaptation to real-world measurement error process without requiring unrealistic assumptions. While we implement the HetGP as the primary statistical model in this work, our framework can accommodate other flexible statistical models for the error process, further extending its applicability (see Section 5). Overall, our approach provides a precise and automated workflow for ODE parameter estimation in settings where measurement error cannot be safely assumed to be i.i.d.\ Gaussian.

\section{Background}
This section provides some fundamental background on foundational concepts that our method is built upon. Specifically, this section summaries essential concepts of Gaussian processes (GPs), heteroscedastic GPs (HetGPs) and Bayesian inference.

\subsection{Gaussian Process}
GPs \citep{NIPS1995_7cce53cf, Rasmussen2006Gaussian} provide a flexible, non-parametric approach to model unknown functions and have been widely used for regression \citep{NIPS1995_7cce53cf}, time-series prediction \citep{Rasmussen2006Gaussian} and the kernel-based regularization method in system identification \citep{PILLONETTO2014657}. A GP defines a prior directly over functions, enabling data-driven surrogate modeling while naturally quantifying predictive uncertainty.

Let $f:\mathbb{R}\to\mathbb{R}$ denote an unknown latent function observed through noisy measurements
\begin{equation}
D=\{(t_i,\tilde y_i)\}_{i=1}^N,\qquad \tilde y_i=f(t_i)+\epsilon_i,
\label{eq:GP_data}
\end{equation}
where $\epsilon_i\sim\mathcal N(0,\sigma_o^2)$ denotes observation error. When $\sigma_o^2$ is constant across inputs, the problem is homoscedastic.

A GP models the latent function \(f(t)\) as
\begin{equation}
    f(t) \sim \mathcal{GP}(\mu(t), k(t, t' \mid \phi)),
\label{eq:gp_prior}
\end{equation}
where \(\mu(t)\) is the mean function representing the expected value of \(f(t)\), and \(k(t, t' \mid \phi)\) is a covariance kernel parametrised by hyperparameters $\phi$. The kernel encodes structural assumptions such as smoothness and characteristic length scales. A common choice for the covariance function is the squared exponential (SE) kernel also called the radial basis function (RBF),
\[
k_{\text{RBF}}(t,t' \mid \phi) = \tau_f^2 \exp\left(-\frac{(t - t')^2}{2\ell^2}\right) \quad \text{with} \quad  \phi = \{\ell, \tau_f\},
\label{eq:rbf}
\]
where \(\ell\) is the length-scale parameter, which determines the distance over which function values are correlated, and \(\tau_f^2\) is the signal variance, controlling the overall variability of the function. Other kernels, such as the Matern family, can also be used to model functions with different smoothness properties.

A key property of GPs is that any finite collection of function values follows a multivariate Gaussian distribution. In particular, evaluating the prior in Eq.~\eqref{eq:gp_prior} at inputs $\{t_i\}_{i=1}^N$ yields
\begin{equation}
\mathbf f = (f(t_1),\ldots,f(t_N))^\top \sim \mathcal N(\boldsymbol\mu,\mathbf K),
\label{eq:gp_finite}
\end{equation}
where $\boldsymbol\mu=(\mu(t_1),\ldots,\mu(t_N))^\top$ and $[\mathbf K]_{ij}=k(t_i,t_j)$.

Under the Gaussian observation model in Eq.~\eqref{eq:GP_data}, the likelihood of the observations conditional on the latent function $\mathbf f$ is
\begin{equation}
p(\tilde{\mathbf y}\mid \mathbf f)=\mathcal N(\mathbf f,\sigma_o^2\mathbf I),
\label{eq:GP_likelihood}
\end{equation}
with $\tilde{\mathbf y}=(\tilde y_1,\ldots,\tilde y_N)^\top$. Combining Eq.~\eqref{eq:GP_likelihood} with the GP prior in Eq.~\eqref{eq:gp_finite} and marginalising the latent vector $\mathbf f$ gives the marginal distribution of the observations,
\begin{equation}
p(\tilde{\mathbf y})=\int p(\tilde{\mathbf y}\mid \mathbf f)\,p(\mathbf f)\,d\mathbf f
=\mathcal N(\boldsymbol\mu,\mathbf K_y),
\label{eq:gp_obs}
\end{equation}
where $\mathbf K_y=\mathbf K+\sigma_o^2\mathbf I$.

The Gaussian form in Eq.~\eqref{eq:gp_obs} provides a closed-form marginal likelihood, enabling hyperparameter learning by maximising  the log marginal likelihood with respect to $\phi$,
\[
\log p(\tilde{\mathbf y}\mid\phi)=
-\frac12(\tilde{\mathbf y}-\boldsymbol\mu)^\top\mathbf K_y^{-1}(\tilde{\mathbf y}-\boldsymbol\mu)
-\frac12\log|\mathbf K_y|
-\frac N2\log 2\pi.
\]
Given the estimated hyperparameters $\hat\phi$, the posterior predictive distribution for the latent function at a new input $t^*$ is given by
\[
f(t^*)\mid\tilde{\mathbf y},\hat\phi\sim\mathcal N(\mu^*(t^*),K^*(t^*)),
\]
with
\[
\mu^*(t^*)=\mu(t^*)+\mathbf k_*^\top\mathbf K_y^{-1}(\tilde{\mathbf y}-\boldsymbol\mu),
\]
\[
K^*(t^*)=k(t^*,t^*)-\mathbf k_*^\top\mathbf K_y^{-1}\mathbf k_*,
\]
where $\mathbf k_*=(k(t_1,t^*),\ldots,k(t_N,t^*))^\top$. The posterior variance $K^*(t^*)$ quantifies uncertainty in the latent function. If predicting a new noisy observation $\tilde y(t^*)$, the observation variance $\sigma_o^2$ is added to $K^*(t^*)$.

\subsection{Heteroscedastic Gaussian Process}
\label{sec:HetGP_background}
The HetGP \citep{Binois_2018,NIPS1997_afe43465, Rasmussen2006Gaussian} extends the standard GP by allowing the observation error to vary with time, providing a more flexible and realistic representation of data exhibiting heteroscedasticity. We assume observations follow
\begin{equation}
\tilde{y}_i = f(t_i) + \epsilon_i, \qquad 
\epsilon_i \sim \mathcal{N}(0,\sigma^2(t_i)),
\label{eq:het_obs}
\end{equation}
where both the latent mean function $f(t)$ and the input-dependent error variance $\sigma^2(t)$ are unknown.

Following \citep{Binois_2018,NIPS1997_afe43465}, we model these quantities as two Gaussian processes:
\begin{equation}
f(t)\sim \mathcal{GP}(\mu_\mu(t),k_\mu(t,t'\mid\phi_\mu)),
\qquad
\log\sigma^2(t)\sim \mathcal{GP}(m_\sigma(t),k_\sigma(t,t'\mid\phi_\sigma)).
\label{eq:hetgp_model}
\end{equation}
Here $\mu_\mu(t)$ and $m_\sigma(t)$ are mean functions, $k_\mu$ and $k_\sigma$ are covariance kernels, and $\phi_\mu,\phi_\sigma$ denote their hyperparameters. The log transformation ensures that the variance remains positive for all inputs.

Let $\mathbf t=(t_1,\dots,t_N)^\top$ and $\tilde{\mathbf y}=(\tilde y_1,\dots,\tilde y_N)^\top$. Marginalising $f$ under the observation model Eq.~\eqref{eq:het_obs} and the GP prior in Eq.~\eqref{eq:hetgp_model} implies
\begin{equation}
\tilde{\mathbf y} \sim \mathcal N(\boldsymbol\mu_\mu,\mathbf K_y),
\qquad
\mathbf K_y=\mathbf K_\mu+\mathrm{diag}(\sigma^2(t_1),\dots,\sigma^2(t_N)),
\label{eq:Ky}
\end{equation}
where $[\mathbf K_\mu]_{ij}=k_\mu(t_i,t_j\mid\phi_\mu)$ and $\boldsymbol\mu_\mu=(\mu_\mu(t_1),\dots,\mu_\mu(t_N))^\top$.

Marginalising the latent function yields an explicit Gaussian likelihood for the observations (Eq.~\eqref{eq:Ky}). This marginal distribution can therefore be used directly for hyperparameter learning. Following standard GP methodology \citep{Binois_2018, Rasmussen2006Gaussian}, we estimate the kernel hyperparameters by maximising the corresponding log marginal likelihood.

The log marginal likelihood of the mean GP, conditional on a given variance function $\sigma^2(t)$, in Eq.~\eqref{eq:Ky} is
\begin{equation}
\log p(\tilde{\mathbf y}\mid \phi_\mu,\sigma^2( t))=
-\frac12(\tilde{\mathbf y}-\boldsymbol\mu_\mu)^\top\mathbf K_y^{-1}(\tilde{\mathbf y}-\boldsymbol\mu_\mu)
-\frac12\log|\mathbf K_y|
-\frac N2\log 2\pi,
\label{eq:lml_mean}
\end{equation}
which depends on the unknown variance function $\sigma^2(t)$ through $\mathbf K_y$.

Since the variance function $\sigma^2(t)$ is unknown, it must be inferred indirectly from the data. Under a Gaussian error assumption, the squared deviation between observations and the posterior mean provides a natural estimate of the local noise level. Therefore, conditional on current estimates of the mean GP, we use squared residuals to construct empirical observations of the variance process.
\begin{equation}
r_i=(\tilde y_i-\hat f(t_i))^2,
\label{eq:residuals}
\end{equation}
and treat $\log r_i$ as empirical noisy observations for $\log\sigma^2(t_i)$, enabling a second GP to be fitted under the log-variance prior in Eq.~\eqref{eq:hetgp_model}. Let $ \mathbf r = ( r_1,\ldots, r_N)^\top$ denote the vector of squared residuals. This gives the variance log marginal likelihood
\begin{equation}
\log p(\log\mathbf r\mid \phi_\sigma)=
-\frac12(\log\mathbf r-\mathbf m_\sigma)^\top\mathbf K_\sigma^{-1}(\log\mathbf r-\mathbf m_\sigma)
-\frac12\log|\mathbf K_\sigma|
-\frac N2\log 2\pi,
\label{eq:lml_var}
\end{equation}
where $[\mathbf K_\sigma]_{ij}=k_\sigma(t_i,t_j\mid\phi_\sigma)$ and $\mathbf m_\sigma=(m_\sigma(t_1),\dots,m_\sigma(t_N))^\top$.

Equations~\eqref{eq:lml_mean} and~\eqref{eq:lml_var} are coupled; Eq.~\eqref{eq:lml_mean} depends on $\sigma^2(t)$ via $\mathbf K_y$, while Eq.~\eqref{eq:lml_var} depends on residuals defined from the posterior mean of $f(t)$ in Eq.~\eqref{eq:residuals}. Therefore, the two likelihoods cannot be optimised independently, motivating the alternating optimisation. Starting from an initial homoscedastic variance estimate, the mean-process hyperparameters $\phi_\mu$ are first obtained by maximising Eq.~\eqref{eq:lml_mean}. Conditioned on this updated mean GP, the squared residuals are computed and treated as empirical observations of the variance process, allowing the variance hyperparameters $\phi_\sigma$ to be estimated by maximising Eq.~\eqref{eq:lml_var}. The resulting posterior mean of the log-variance GP yields updated variance estimates $\hat{\sigma}^2(t_i)$, which are then fed back into Eq.~\eqref{eq:lml_mean} for the next iteration. This alternating optimisation scheme updates the mean and variance hyperparameters in turn, holding one fixed while optimising the other, until convergence of both hyperparameter sets and variance estimates, effectively performing joint maximum likelihood estimation of the coupled mean and variance processes.

In this work we use constant zero mean functions, $\mu_\mu(t)= 0$ and $m_\sigma(t)= 0$, which is common in GP modeling after centering and scaling the observations \citep{Binois_2018, NIPS1997_afe43465} . This avoids introducing additional mean parameters while retaining flexibility through the covariance structure in Eq.~\eqref{eq:hetgp_model} (see Section~\ref{sec:step1}).  If strong global trends remain, parametric mean terms can be incorporated without changing the framework.

Algorithm~\ref{alg:HetGP} summarises this iterative procedure.

\begin{algorithm}[H]
\caption{HetGP algorithm for learning a time-varying observation variance.}
\label{alg:HetGP}
\begin{algorithmic}[1]
\REQUIRE Inputs $\mathbf t=(t_1,\dots,t_N)^\top$, observations $\tilde{\mathbf y}=(\tilde y_1,\dots,\tilde y_N)^\top$, kernels $k_\mu(\cdot,\cdot|\phi_\mu)$ and $k_\sigma(\cdot,\cdot|\phi_\sigma)$, tolerance $\epsilon$, initial variance $\sigma_0^2$, maximum iterations $J_{\max}$
\STATE Initialize $\sigma_{(0)}^2(t_i) = \sigma_0^2$ for $i=1,\dots,N$
\FOR{$j = 1,2,\dots,J_{\max}$}
    \STATE \textbf{Mean step:}
    \STATE \quad Form $\mathbf{K}_y = \mathbf{K}_\mu(\phi_\mu^{(j-1)}) + \operatorname{diag}(\sigma^2_{(j-1)}(t_1), \dots, \sigma^2_{(j-1)}(t_N))$
    \STATE \quad Maximize $\log p(\tilde{\mathbf{y}} \mid \phi_\mu, \sigma^2_{(j-1)}(\mathbf t))$ to obtain $\phi_\mu^{(j)}$
    \STATE \quad Compute posterior mean $\hat f^{(j)}(t_i)$
    \STATE \textbf{Residuals:}
    \STATE \quad Compute squared residuals $r_i=(\tilde y_i-\hat f^{(j)}(t_i))^2$
    \STATE \quad Set $\mathbf{z} = \log \mathbf{r} = \log(r_1,\dots,r_N)^\top$
    \STATE \textbf{Variance-step:}
    \STATE \quad Maximize $\log p(\mathbf{z} \mid \phi_\sigma)$ to obtain $\phi_\sigma^{(j)}$
    \STATE \quad Compute posterior mean $\hat\mu_\sigma^{(j)}(t_i)$
    \STATE \quad Update $\sigma_{(j)}^2(t_i)=\exp(\hat\mu_\sigma^{(j)}(t_i))$
    \STATE \textbf{Convergence check:}
    \IF{$\max \Big(|\phi_\mu^{(j)} - \phi_\mu^{(j-1)}|, |\phi_\sigma^{(j)} - \phi_\sigma^{(j-1)}|, \max_i |\sigma^2_{(j)}(t_i) - \sigma^2_{(j-1)}(t_i)|\Big) < \epsilon$}
    \STATE \textbf{break}
    \ENDIF
    \ENDFOR

\RETURN
$\hat\phi_\mu,\hat\phi_\sigma,\{\hat\sigma^2(t_i)\}_{i=1}^N$
\end{algorithmic}
\end{algorithm}

\subsection{Bayesian Inference}
Bayesian inference provides a general framework for parameter estimation under uncertainty. Let $\theta \in \mathbb{R}^m$ denote the parameter vector of an ODE model, and let $\tilde{\mathbf y} = (\tilde y(t_1),\dots,\tilde y(t_N))$ denote the observed data at times $\mathbf t = (t_1,\dots,t_N)$. Given a prior distribution $p(\theta)$ and likelihood $p(\tilde{\mathbf{y}}\mid\theta)$, Bayes’ theorem yields \citep{Gelman_2013}
\[
p(\theta \mid \tilde{\mathbf y}) 
= \frac{p(\tilde{\mathbf y} \mid \theta) p(\theta)}
       {p(\tilde{\mathbf y})},
\]
where the normalising constant $p(\tilde{\mathbf y})$ is typically intractable, implying that direct independent sampling from the posterior is generally not feasible. Statistical inference therefore requires sampling-based methods such as Markov chain Monte Carlo (MCMC) \citep{Robert_2011}. MCMC constructs a Markov chain whose stationary distribution coincides with the posterior, enabling sampling when independent draws are infeasible. The initial portion of the chain is discarded as a burn-in phase to reduce the influence of starting values and allow the chain to approach its stationary distribution. After running the chain for a sufficiently long burn-in period, the generated samples approximate draws from the posterior without requiring evaluation of the intractable normalising constant in Bayes’ theorem.

We use the Metropolis-Hastings (MH) \citep{10.1063/1.1699114} algorithm, a widely used MCMC technique, to sample from the posterior distribution of the parameters. Starting from an initial value $\theta^{0}$, at iteration $i$ a candidate state $\theta'$ is proposed from a proposal distribution $q(\theta' \mid \theta^{i-1})$, typically taken as a Gaussian random walk:
\[
\theta' \sim \mathcal{N}(\theta^{i-1}, \Lambda),
\]
where $\Lambda$ is the proposal covariance matrix that controls the step size and correlation structure of the proposed moves. Choosing an appropriate $\Lambda$ is important to balance exploration of the parameter space and acceptance rate; theoretical results suggest an optimal acceptance rate around 0.234 for high-dimensional problems \citep{c56b4e38-c4b3-34ac-8e24-87699314559b}.

Mathematically, given a current state $\theta^{i-1}$, a proposal distribution $q(\theta' | \theta^{i-1})$, and the posterior density $p(\theta| \tilde{\mathbf y})$, the acceptance probability for moving to the new state $\theta'$ is calculated as:
\[  \alpha(\theta' , \theta^{i-1}) = \min \left(1, \frac{q(\theta^{i-1} | \theta') p(\theta'|\tilde{\mathbf y})}{q({\theta'} | \theta^{i-1}) p(\theta^{i-1}|\tilde{\mathbf y})}\right).
\]
Since the posterior is required only up to a proportionality constant, the intractable normalising constant cancels in this acceptance ratio. With probability $\alpha(\theta' , \theta^{i-1})$, the proposal is accepted, $\theta^i \leftarrow \theta'$, otherwise the chain remains at the current state, $\theta^i \leftarrow \theta^{i-1}$. This iterative procedure produces a Markov chain whose stationary distribution corresponds to the target posterior $p(\theta \mid \tilde{\mathbf y})$. The pseudo-code of the MH algorithm is shown in Algorithm~\ref{MH}.

\begin{algorithm}[H]
\caption{Metropolis-Hastings Algorithm}
\label{MH}
\begin{algorithmic}[1]
\STATE{ Initial state $\theta^{(0)}$}
\FOR{$i = 1, 2, ...$}
    \STATE{ Propose: $\theta' \sim q(\cdot | \theta^{i-1})$}
    \STATE{ Acceptance Probability: $\alpha(\theta' , \theta^{i-1}) = \min \left(1, \frac{q(\theta^{i-1} | \theta') p(\theta'|\tilde{\mathbf y})}{q({\theta'} | \theta^{i-1}) p(\theta^{i-1}|\tilde{\mathbf y})}\right)$}
    \STATE{ $u \sim \text{Uniform}(0, 1)$}
    \IF{$u < \alpha(\theta' , \theta^{i-1})$}
        \STATE{ Accept the proposal: $\theta^{i} \leftarrow \theta'$}
    \ELSE
        \STATE {Reject the proposal: $\theta^{i} \leftarrow \theta^{i-1}$}
    \ENDIF
\ENDFOR
\end{algorithmic}
\end{algorithm}

\section{A Two-step Approach for Dealing with Heteroscedastic Error}
\label{our_method}
Specifying appropriate error models for mechanistic ODE systems is challenging, and inadequate assumptions about the error structure can lead to biased parameter estimates and inaccurate uncertainty quantification. In contrast, purely data-driven approaches such as heteroscedastic Gaussian processes can flexibly capture variability in observations but lack mechanistic interpretability. To address this issue, we propose a two-step framework that combines flexible statistical modeling of measurement error with Bayesian inference for mechanistic ODE models. The key idea is to separate estimation of the observation error process from inference of the dynamic parameters. In the first step, the structure of the measurement error process is learned directly from the data using HetGP, thereby separating stochastic variability from the underlying dynamics (Section~\ref{sec:step1}). In the second step, the estimated error process is treated as known and incorporated into the Bayesian inference procedure via the likelihood function, enabling principled estimation of the ODE parameters (Section~\ref{sec:step2}). This separation allows stochastic variability to be modeled nonparametrically, while preserving interpretability of the mechanistic model and decreasing the bias caused by measurement error when characterizing system dynamics.

\subsection{Step 1: Estimation of the Measurement Error Process}
\label{sec:step1}

Let $\{\tilde{y}(t_i)\}_{i=1}^N$ denote noisy observations of an ODE system at times $\{t_i\}_{i=1}^N$, as defined in Equations~\eqref{eq:ODE} and \eqref{eq:obs_model}. In our framework, the measurement error $\epsilon(t_i)$ may be homoscedastic, heteroscedastic, or even correlated. In this work we focus on time-varying measurement error, extensions to autocorrelated errors are discussed in Section~\ref{sec:discussion}.

To estimate the input-dependent measurement error, we employ HetGP regression as described in Section~\ref{sec:HetGP_background}. Specifically, following \citep{Binois_2018}, we model the latent mean function and the log-variance as two coupled Gaussian processes,
\[
    f(t) \sim \mathcal{GP}\big(0, k_\mu(t,t'\mid \phi_{\mu})\big), \quad
    \log \sigma^2(t) \sim \mathcal{GP}\big(0, k_\sigma(t, t' \mid \phi_\sigma)\big),
\]
where zero mean functions are adopted in accordance with standard practice after centering the data \citep{Binois_2018}. The covariance kernels $k_\mu$ and $k_\sigma$ control the smoothness of the latent mean and variance processes, respectively.

Inference proceeds by maximising the coupled log marginal likelihoods in Equations~\eqref{eq:lml_mean} and~\eqref{eq:lml_var} via the alternating optimisation scheme summarised in Algorithm~\ref{alg:HetGP}. At each iteration, the mean GP is updated conditional on current variance estimates, squared residuals are computed according to Eq.~\eqref{eq:residuals}, and a second GP is fitted to the log-residuals to update the variance process. The posterior mean of the variance GP provides updated estimates
\[
\hat\sigma^2(t_i)=\exp\big(\hat\mu_\sigma(t_i)\big),
\]
which are then fed back into the mean GP in the next iteration.

In practice, this procedure is implemented using the \texttt{mleHetGP} function from the \texttt{HetGP} R package. The output of Step 1 consists of fitted hyperparameters $(\hat\phi_\mu,\hat\phi_\sigma)$ and estimated time-dependent observation variances $\hat\sigma^2(t_i)$ at each observation time. These variance estimates are treated as empirical approximations of the measurement error process and are subsequently used in the Bayesian inference stage.

In addition, the implementation allows automatic comparison between heteroscedastic and homoscedastic noise models through the argument \texttt{checkHom} in \texttt{mleHetGP}. When \texttt{checkHom = TRUE}, the procedure evaluates the maximum marginal log-likelihood under both specifications and returns the homoscedastic model only if it achieves a higher value. This option therefore allows comparison between constant and input-dependent noise models based on their marginal likelihoods.

\subsection{Step 2: Bayesian Inference for ODE Parameters}
\label{sec:step2}
Given the estimated variance function $\hat{\sigma}^2(t_i)$ obtained in Step~1, we define a heteroscedastic Gaussian observation model for the ODE parameters consistent with Eq.~\eqref{eq:obs_model}:
\begin{equation}
\log p(\tilde{\mathbf y} \mid \theta, \hat{\sigma}^2(\mathbf{t})) 
= -\frac{1}{2} \sum_{i=1}^N \left[
\log(2\pi \hat{\sigma}^2(t_i)) + 
\frac{(\tilde{y}(t_i) - g(y(t_i;\theta)))^2}{\hat{\sigma}^2(t_i)}
\right].
\label{eq:hetero_likelihood}
\end{equation}
where $y(t_i;\theta)$ denotes the solution of the ODE system in Eq.~\eqref{eq:ODE} evaluated at time $t_i$.

For comparison, under the standard homoscedastic assumption with constant variance $\sigma^2$, the likelihood takes the form
\begin{equation}
\log p(\tilde{\mathbf{y}} \mid \theta, \sigma^2) 
= -\frac{1}{2} \sum_{i=1}^N \left[
\log(2\pi \sigma^2) + 
\frac{(\tilde{y}(t_i) - g(y(t_i;\theta)))^2}{\sigma^2}
\right].
\label{eq:homo_likelihood}
\end{equation}
Thus, the heteroscedastic formulation in Eq.~\eqref{eq:hetero_likelihood} generalizes the formulation in Eq.~\eqref{eq:homo_likelihood} by allowing the variance to vary over time. Posterior inference then combines this likelihood with prior distributions $p(\theta)$ on the ODE parameters:
\[
    p(\theta \mid \tilde{\mathbf{y}}) \propto 
    p(\tilde{\mathbf{y}} \mid \theta, \hat{\sigma}^2(\mathbf{t})) p(\theta).
\]
We draw samples from this posterior using a random-walk Metropolis-Hastings Markov chain Monte Carlo algorithm (Algorithm~\ref{MH}). For each proposed parameter vector $\theta'$, the predicted trajectories $y(t_i;\theta')$ are obtained using a numerical solution or an analytical solution when available and substituted into Eq.~\eqref{eq:hetero_likelihood}. Convergence of the sampler is monitored using the potential scale reduction factor $\hat{R}$ \citep{Gelman_2013}, with all parameters achieving $\hat{R} < 1.1$ in our example. Traceplots were further examined to qualitatively assess if there was any evidence of a lack of convergence.

This two-step approach avoids joint optimisation over both components while still accounting for heteroscedastic uncertainty, leading to improved robustness and computational efficiency. The overall workflow is summarised in Algorithm~\ref{alg:Method}.
\begin{algorithm}[H]
\caption{Two-step Bayesian inference with HetGP error estimation}
\label{alg:Method}
\begin{algorithmic}[1]
\STATE \textbf{Input:} Observed data $\mathbf{y}$, ODE model $h(y,t;\theta)$, prior distribution $p(\theta)$
\STATE \textbf{Step 1: Time-varying error estimation (HetGP)}
  \STATE \quad Model latent mean and log-variance with Gaussian processes:
  
  \quad $f(t) \sim \mathcal{GP}(0, k_\mu(t,t'\mid \phi_\mu)), \quad \log \sigma^2(t) \sim \mathcal{GP}(0, k_\sigma(t,t'\mid \phi_\sigma))$
  \STATE \quad Fit HetGP model (Algorithm~\ref{alg:HetGP}) to obtain time-varying error $\hat{\sigma}^2(t_i)$
\STATE \textbf{Step 2: Bayesian inference for ODE parameters}
  \STATE \quad Construct heteroscedastic likelihood using learned $\hat\sigma^2(t_i)$
  \STATE \quad Form posterior distribution $p(\theta \mid \mathbf{y}) \propto p(\mathbf{y}\mid\theta,\hat\sigma^2_\mathbf{t})\,p(\theta)$
  \STATE \quad Draw posterior samples using the MH algorithm (see Algorithm~\ref{MH})
\STATE \textbf{Output:} Posterior samples of $\theta$
\end{algorithmic}
\end{algorithm}

\section{Results}
We evaluated the proposed heteroscedastic inference framework across three case studies of increasing complexity. The first is a controlled simulation study based on logistic growth with time-varying measurement error. This setting allows systematic assessment of how error misspecification influences posterior inference as the number of observations increases. Posterior accuracy is quantified using the maximum mean discrepancy (MMD), enabling comparison of homoscedastic and heteroscedastic models.

The second and third examples involve real-world ecological and epidemiological data, respectively. 
These applications introduce increasing structural and computational complexity, including analytically solvable and numerically integrated ODE systems, as well as pronounced temporal variability in the observation process. 
In all cases, we compare the proposed heteroscedastic framework with a standard homoscedastic Gaussian error model.

\subsection{Example 1: A Simulation Study of Logistic Growth}
We first consider a simulation study using the logistic growth model, a canonical nonlinear ODE used in many modeling applications of biological and ecological science to describe density-dependent population dynamics \citep{Maclaren2017, Murphy2022}, epidemiological spread \citep{Alahmadi2020, Browning2020}, and ecosystem recovery \citep{Warne_2025,https://doi.org/10.1111/1365-2664.14039}. Its simplicity, combined with its capacity to represent density-dependent constraints on growth, makes it a natural choice for assessing the performance of our inference framework. The logistic growth model is defined by:
\begin{equation}
\frac{\textrm{d}C}{\textrm{d}t} = rC\left(1 - \frac{C}{K}\right),
\label{eq:logistic_growth}
\end{equation}
where $C$ denotes the population size as a function of time $t$, $r$ is the intrinsic per capita growth rate in the absence of constraints, $K$ is the carrying capacity which is the maximum population size that the environment can sustain. Given $C(0)=C_0$ the analytical solution of Eq.~\eqref{eq:logistic_growth} is
\[
C(t) \;=\; \frac{K\,C_0}{(K-C_0)e^{-r t} + C_0}.
\label{eq:logistic_solution}
\]

This solution exhibits sigmoid growth, characterised by an initial exponential phase followed by saturation toward the carrying capacity, with $C(t) \to K$ as $t \to \infty$.
The purpose of this example is to assess, using an extensive simulation study, how different error assumptions affect posterior inference as the number of observations increases. Datasets were generated by solving the logistic ODE with parameters $r=0.0025$, $K=80$, and $C_0=0.7$ across $t = \{0,\dots,4000\}$, where $t$ is measured in days. These parameter values were chosen to produce a gradual sigmoid growth trajectory over a realistic temporal scale, allowing the effects of observation error and sampling density to be clearly distinguished. Observations were generated under a heteroscedastic Gaussian error model consistent with Eq.~\eqref{eq:obs_model},
\[
C_{\mathrm{obs}}(t_i) = C(t_i) + \epsilon(t_i), 
\qquad \epsilon(t_i)\sim \mathcal{N}\!\big(0,\sigma_t^2\big),
\]
where the time-dependent standard deviation increases with time:
\begin{equation}
    \sigma_t = \sigma \sqrt{\frac{t}{100}},
    \label{Eq:known_sigma}
\end{equation}
where $\sigma = 2.3$ is the true error used to generate the synthetic datasets.
For each of six dataset sizes ($n = 10,20,50,100,500,1000$), we generated 25 datasets by sampling at equally spaced intervals, yielding 150 datasets in total.

We performed Bayesian inference for $(r,K,\sigma)$ using Metropolis-Hastings MCMC. Uniform priors were assigned within their respective bounds, $r\in[0.001,1]$, $K\in[10,100]$, and $\sigma\in[1,20]$. To sample in an unconstrained space, each bounded parameter was mapped to $\mathbb{R}$ via a logit transform. Each chain was run for 10{,}000 iterations with the first 1000 discarded as burn-in. We compare three likelihood specifications: (i) a homoscedastic model $C_{\mathrm{obs}}(t_i)\sim \mathcal{N}(C(t_i),\sigma^2)$; (ii) our proposed heteroscedastic model, where $\sigma^2$ is replaced by the HetGP estimate $\hat\sigma^2(t_i)$ and used in the likelihood of Eq.~\eqref{eq:hetero_likelihood}; and (iii) using the true $\sigma_t$ from Eq.~\eqref{Eq:known_sigma} providing a benchmark for evaluating the performance of the other two models.

For each simulated dataset, posterior samples of the ODE parameters were obtained under each likelihood formulation, resulting in three sets of posterior samples, $\{\Theta_{\text{homoscedastic}}^{(s)}\}_{s=1}^S$, $\{\Theta_{\text{our approach}}^{(s)}\}_{s=1}^S$, $\{\Theta_{\text{true}}^{(s)}\}_{s=1}^S$ , where each set represents draws from the corresponding posterior distribution and $S$ denotes the number of posterior samples. To quantify agreement between the approximate posteriors and the benchmark posterior, we computed the maximum mean discrepancy (MMD) \citep{JMLR:v13:gretton12a} between posterior samples. Specifically, we evaluate
\[
\mathrm{MMD}(\{\Theta_{\text{homoscedastic}}^{(s)}\}, \{\Theta_{\text{true}}^{(s)}\}) 
\quad \text{and} \quad
\mathrm{MMD}(\{\Theta_{\text{our approach}}^{(s)}\}, \{\Theta_{\text{true}}^{(s)}\}).
\]
Given two sets of samples, $X = \{x_1, \dots, x_n\}$ drawn from distribution $P$ and $Z = \{z_1, \dots, z_m\}$ drawn from distribution $Q$, the squared MMD is defined as:
\[
\mathrm{MMD}^2(P,Q) =
\frac{1}{n^2} \sum_{i=1}^{n} \sum_{j=1}^{n} k(x_i, x_j) +
\frac{1}{m^2} \sum_{i=1}^{m} \sum_{j=1}^{m} k(z_i, z_j) -
\frac{2}{nm} \sum_{i=1}^{n} \sum_{j=1}^{m} k(x_i, z_j),
\]
where $k(\cdot, \cdot)$ is a positive-definite kernel function. We adopt the Gaussian radial basis function (RBF) kernel:
\[
k(x, z) = \exp \left( -\frac{\|x - z\|^2}{2 \lambda^2} \right).
\]
The bandwidth parameter $\lambda$ is set using the median heuristic, a widely used strategy in kernel two-sample testing \citep{JMLR:v13:gretton12a}. Specifically, we compute all pairwise distances between samples from both distributions, take their median, and use this value as $\lambda$. This approach balances sensitivity to both local and global structure, has strong empirical performance, and is the default in many MMD implementations \citep{JMLR:v13:gretton12a}.

\begin{figure}[ht!]
\centering
\includegraphics[width=1\textwidth]{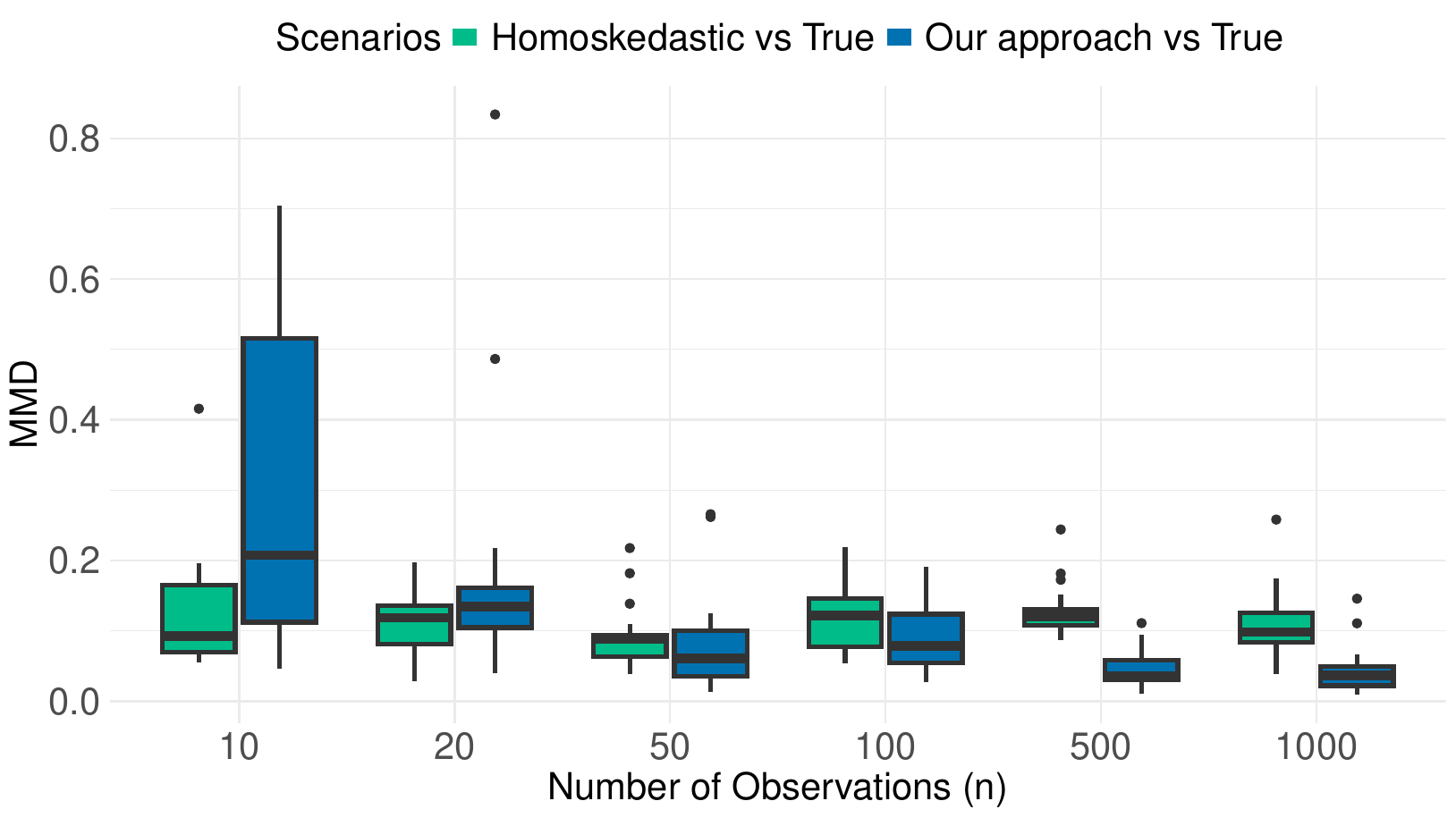}
\captionsetup{
    width=0.9\textwidth,
    font=footnotesize,  
    labelfont=bf,      
    justification=justified,
    skip=6pt         
}
\caption{MMD comparison between the posterior samples obtained from the homoscedastic model and our approach relative to the true posterior. For each dataset size ($n = 10, 20, 50, 100, 500, 1000$), 25 datasets under the same fixed parameter values were generated by sampling the ODE solution at equally spaced time points. This produces 25 MMD values for each combination of method and dataset size, and these 25 MMD values are represented as a boxplot. Lower MMD values indicate closer alignment with the true posterior. Note that for visualization clarity, boxplots for our approach are slightly offset to the right and those for the homoscedastic model slightly offset to the left of each dataset size shown on the x-axis.}
\label{fig:mmd_boxplot}
\end{figure}

Figure~\ref{fig:mmd_boxplot} summarises MMD values across all datasets. Our HetGP-based approach predominantly produces approximate posterior samples closer to the true posterior compared to the homoscedastic model, except for very small sample sizes. Furthermore, for our approach the MMD decreases with sample size, and our approach leads to a larger relative improvement compared with the homoscedastic model. These results demonstrate that ignoring heteroscedasticity leads to systematically biased posterior approximation, particularly as sample size increases.

For all simulated datasets, the automatic model selection procedure in \texttt{mleHetGP} (\texttt{checkHom = TRUE}) returned the same results as the forced heteroscedastic fit (\texttt{checkHom = FALSE}), indicating that the heteroscedastic model was preferred under the marginal likelihood criterion.

\FloatBarrier

\subsection{Example 2: Coral Reef Recovery}

The second example applies the methodology to real ecological data, focusing on the recovery of coral cover following a disturbance. This example provides an ecological growth system in which uncertainty quantification is directly relevant to forecasting. The model represents a relatively low-dimensional nonlinear ODE system, allowing us to examine how heteroscedastic error influences inference in a practical setting where the underlying ODE structure is available.
The coral cover time-series data were obtained from benthic surveys conducted by the Australian Institute of Marine Sciences (AIMS) Long Term Monitoring Program (LTMP) and Marine Monitoring Program (MMP) between 1992 and 2020 \citep{Warne_2025}. For each time point the dataset provides the mean percentage hard coral cover (C) and an associated standard deviation (C$_{sd}$) \citep{Warne_2025}. 
This allows us to incorporate the empirically reported measurement uncertainty directly into our model, preserving the heteroscedastic structure observed in the monitoring data. Specifically, for each time point $t_i$, $J=5$ synthetic observations were generated according to:
\[
y_{ij} \sim \mathcal{N}\!\big(C(t_i), C_{sd}(t_i)\big), \quad y_{ij}>0 ,\quad j=1,\dots,J,
\]
where $y_i$ denotes the synthetic observations. This reconstruction approximates the underlying transect-level variability while preserving the heteroscedastic measurement uncertainty.

The dynamics were modeled using the generalized logistic (Richards') growth form \citep{10.1093/jxb/10.2.290,SIMPSON2022110998, TSOULARIS200221}:
\begin{equation}
\label{eq:coral_ODE}
\frac{\mathrm{d}C(t)}{\mathrm{d}t} = \frac{\alpha}{\gamma} C(t) \left(1 - \frac{C(t)}{K}\right)^\gamma \quad t > t_0,
\end{equation}
where $C(t) \in (0,100]$ (\% area) is the hard coral cover at time $t$ in years, $\alpha > 0$ (1/year) is the intrinsic growth rate governing the maximal rate of coral cover increase in the absence of competition, $K \in (0,100]$ (\% area) is the long-term carrying capacity representing the maximum feasible coral cover, and $\gamma > 0$ is a non-dimensional generalisation parameter that controls the shape of the sigmoid growth curve, regulating the curvature of the trajectory as $C(t)$ approaches $K$.

An analytical solution of Eq.~\eqref{eq:coral_ODE} exists and takes the form:
\[
C(t) = K \left[ 1 + \left( \left(\frac{K}{C_0}\right)^\gamma - 1 \right) e^{-\alpha t} \right]^{-1/\gamma}.
\]

\begin{figure}[ht!]
\centering
\includegraphics[width=0.85\textwidth]{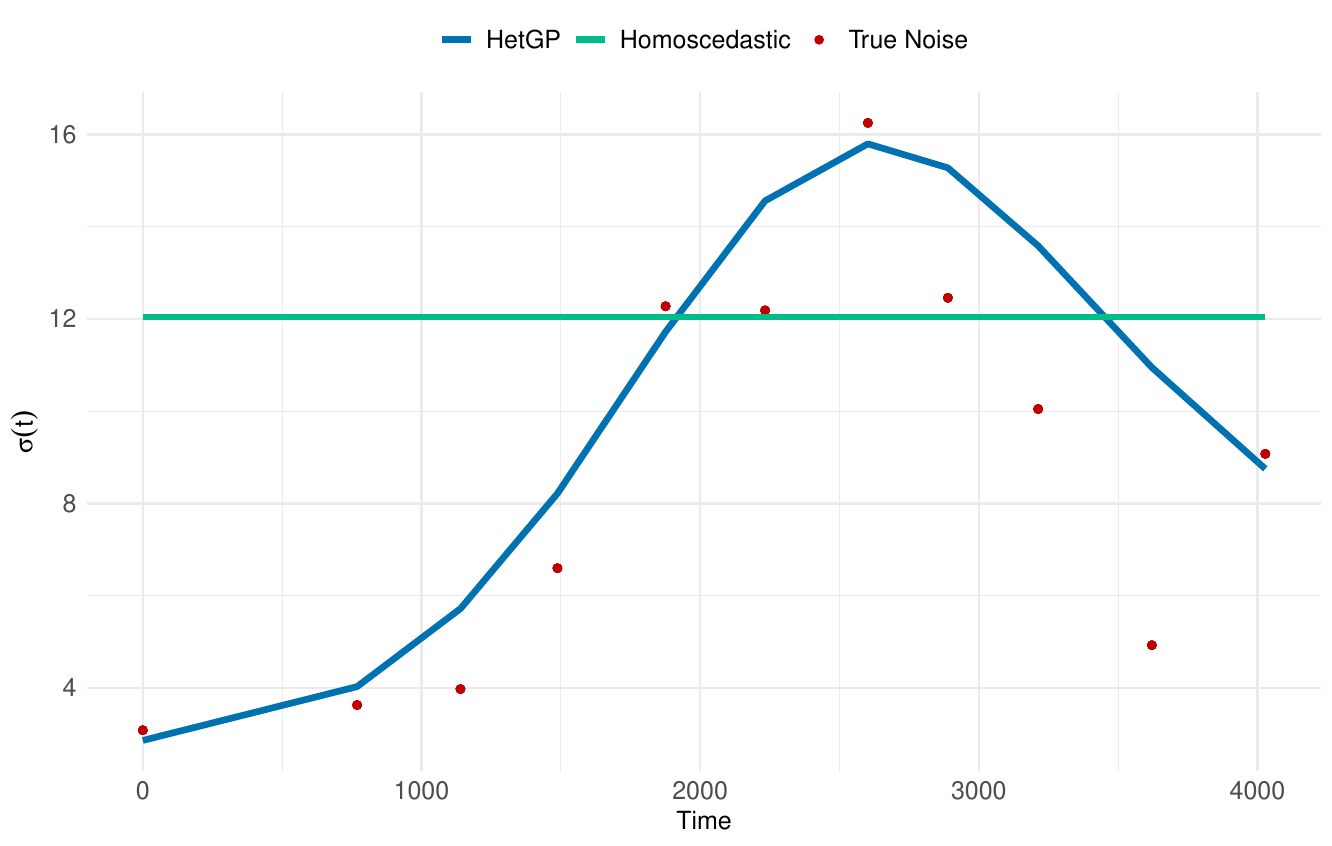}
\captionsetup{
    width=0.9\textwidth,
    font=footnotesize,  
    labelfont=bf,      
    justification=justified,
    skip=6pt         
}
\caption{Comparison of the estimated error over time, showing the HetGP fit (blue), the homoscedastic model (green), and the true error level used in the dataset (red points).}
\label{fig:coral_sigma}
\end{figure}

The estimated time-varying error is shown in Figure~\ref{fig:coral_sigma}. The HetGP estimate varies systematically over time, capturing the increasing and decreasing variability present in data. The automatic selection option in \texttt{mleHetGP} produced estimates identical to the forced heteroscedastic fit, again indicating that the heteroscedastic specification achieved a higher marginal likelihood than the homoscedastic alternative.

We perform MCMC sampling for the posterior distributions of $\alpha$, $\gamma$, and $K$, using logit and log transformations as appropriate for bounded and positive parameters. We assigned priors to the model parameters as $\alpha \sim \text{Uniform}(0,1)$, $K \sim \text{Uniform}(0,100)$, and $\gamma \sim \text{Exponential}(\lambda)$ where $\lambda=1$. To enable MCMC sampling in an unbounded space, $\alpha$ and $K$ were transformed using a logit function over their respective bounds, and $\gamma$ was log-transformed. Posterior samples were drawn using MH algorithm for $10{,}000$ iterations, with the first $2{,}000$ iterations were discarded as burn-in.

\begin{figure}[ht!]
\centering
\includegraphics[width=1\textwidth]{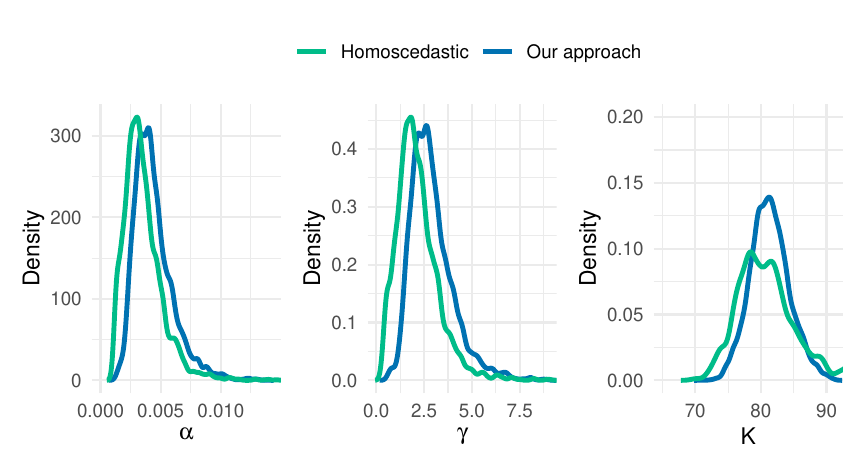}
\captionsetup{
    width=0.9\textwidth,
    font=footnotesize,  
    labelfont=bf,      
    justification=justified,
    skip=6pt         
}
\caption{Kernel Density Estimation (KDE) of the marginal posterior distributions for parameters $\alpha$ (growth rate), $\gamma$ (shape parameter), and $K$ (carrying capacity) under different error models. The blue curves correspond to our approach and the green curves to the homoscedastic model.}
\label{fig:2_3}
\end{figure}

Kernel density estimation were used to obtain smooth representations of the marginal posterior distributions based on the MCMC samples. Since the univariate posterior distributions of $\alpha$, $\gamma$, and $K$ are largly similar under both homoscedastic and heteroscedastic assumptions (Figure~\ref{fig:2_3}), we further examine the joint posterior structure.

\begin{figure}[ht!]
\centering
\includegraphics[width=1\textwidth]{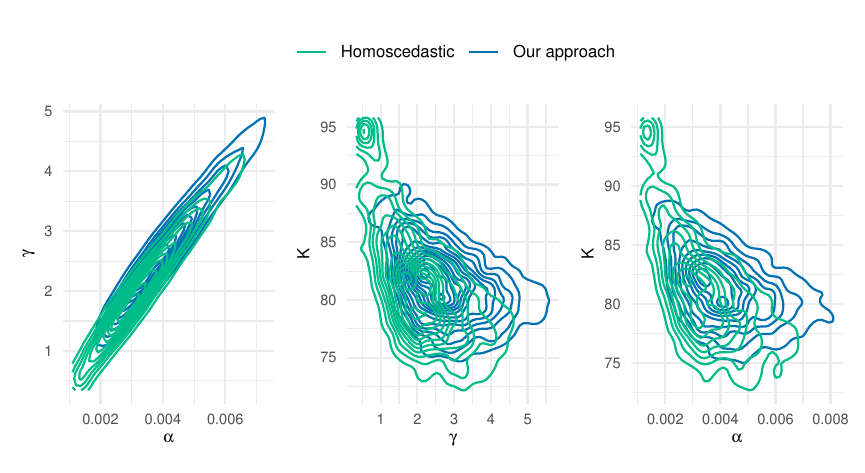}
\captionsetup{
    width=0.9\textwidth,
    font=footnotesize,  
    labelfont=bf,      
    justification=justified,
    skip=6pt         
}
\caption{Bivariate posterior distributions for parameter pairs $(\alpha, \gamma)$, $(\alpha, K)$, $(\gamma, K)$ comparing our approach (blue) and homoscedastic (green) model.}
\label{fig:bivariate_posterior}
\end{figure}

We explored the bivariate posterior distributions of the parameter pairs $(\alpha, \gamma)$, $(\alpha, K)$, and $(\gamma, K)$ (Figure~\ref{fig:bivariate_posterior}). Although the overall dependence patterns are similar across models, the heteroscedastic framework produces slightly wider joint uncertainty regions and modest shifts in the correlation structure between parameters. This implies that comparable marginal distributions do not necessarily guarantee similar joint behaviour, as the assumed error structure can modify the shape of the posterior surface and the interaction between parameters.

\begin{figure}[ht!]
\centering
\includegraphics[width=1\textwidth]{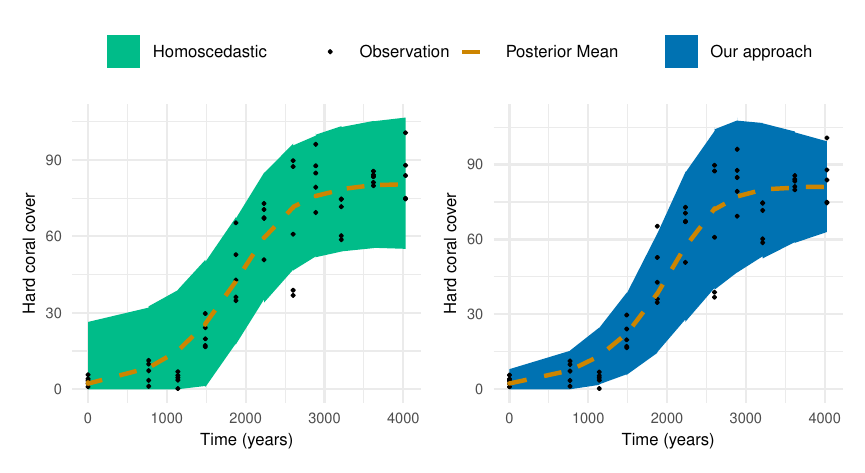}
\captionsetup{
    width=0.9\textwidth,
    font=footnotesize,  
    labelfont=bf,      
    justification=justified,
    skip=6pt         
}
\caption{Comparison of posterior predictive intervals in Coral cover time series data. The left plot shows a symmetric credible interval across all time points, while the right plot shows heteroscedastic interval that adapts to varying error levels.}
\label{fig:coral_fit}
\end{figure}

The posterior predictive intervals for both homoscedastic and heteroscedastic error models are shown in Figure~\ref{fig:coral_fit}. The homoscedastic model assumes constant variance and consequently produces predictive intervals with nearly uniform width across time. In contrast, our approach allows the observation variance to
vary with time, leading to predictive intervals that better reflect the time-varying uncertainty in the observations.

\FloatBarrier

\subsection{Example 3: The 1984 Measles Epidemic}
The final example evaluates the method on epidemiological data, using weekly measles incidence in England and Wales \citep{Dattner2015AMI, 2020_Huang} . This example extends the analysis to an infectious disease transmission model requiring numerical integration. 
Compared to the previous examples, epidemic dynamics involve sharper temporal changes in infection levels, providing a more demanding test of heteroscedastic modeling in practice. The underlying dynamics were represented by a susceptible–infected–recovered (SIR) framework, 
\begin{equation}
\begin{aligned}
\frac{\mathrm{d}S}{\mathrm{d}t} = -\beta S(t) I(t), \\
\frac{\mathrm{d}I}{\mathrm{d}t}=\beta S(t) I(t)-\delta I(t),
\end{aligned}
\label{Eq.measles}
\end{equation}
where $S(t)$ and $I(t)$ denote the number of susceptible and infected individuals at time $t$ respectively, $\beta$ is the infection rate and $\delta$ is the recovery rate. We follow \citep{Dattner2015AMI,2020_Huang} and focus on the 1948-1949 measles epidemic data, assumed to begin around day 254 (early September), where social mixing due to school term initiates transmission cycles. In line with \citep{Dattner2015AMI,2020_Huang} we focus on a subset of 53 weeks of reported cases starting from this point. Observed measles case counts display pronounced seasonal and demographic variability, which translates into non-constant variance across time. 
We integrate the first equation  of the system in Eq.~\eqref{Eq.measles} to obtain
\[
S(t) = S_0 \exp\Big(-\int_0^t \beta I(s)\,\mathrm{d}s \Big).
\]
Substituting this into the equation for $I(t)$ gives
\[
\frac{\mathrm{d}I(t)}{\mathrm{d}t} = \Big(\beta S_0 \exp\Bigl(-\int_0^t \beta I(s)\,\mathrm{d}s\Bigr) - \delta \Big) I(t).
\]
Following \citep{Dattner2015AMI}, we assume that an individual recovers in 5 days, which corresponds to $\delta = 7/5$ given that the data are reported weekly. With $\delta$ fixed, the parameters are inferred from the data. Parameters $\theta = (\beta,S_0,I_0)$ considered to be positive ($\theta_i>0$), were estimated on the log scale to ensure positivity, with weakly informative independent Gaussian priors centred at zero with unit variance on the transformed scale. We sample posteriors using MH algorithm and chains were run for $10{,}000$ iterations, and the first $2{,}000$ iterations were discarded as burn-in.
\begin{figure}[ht!]
\centering
\includegraphics[width=0.85\textwidth]{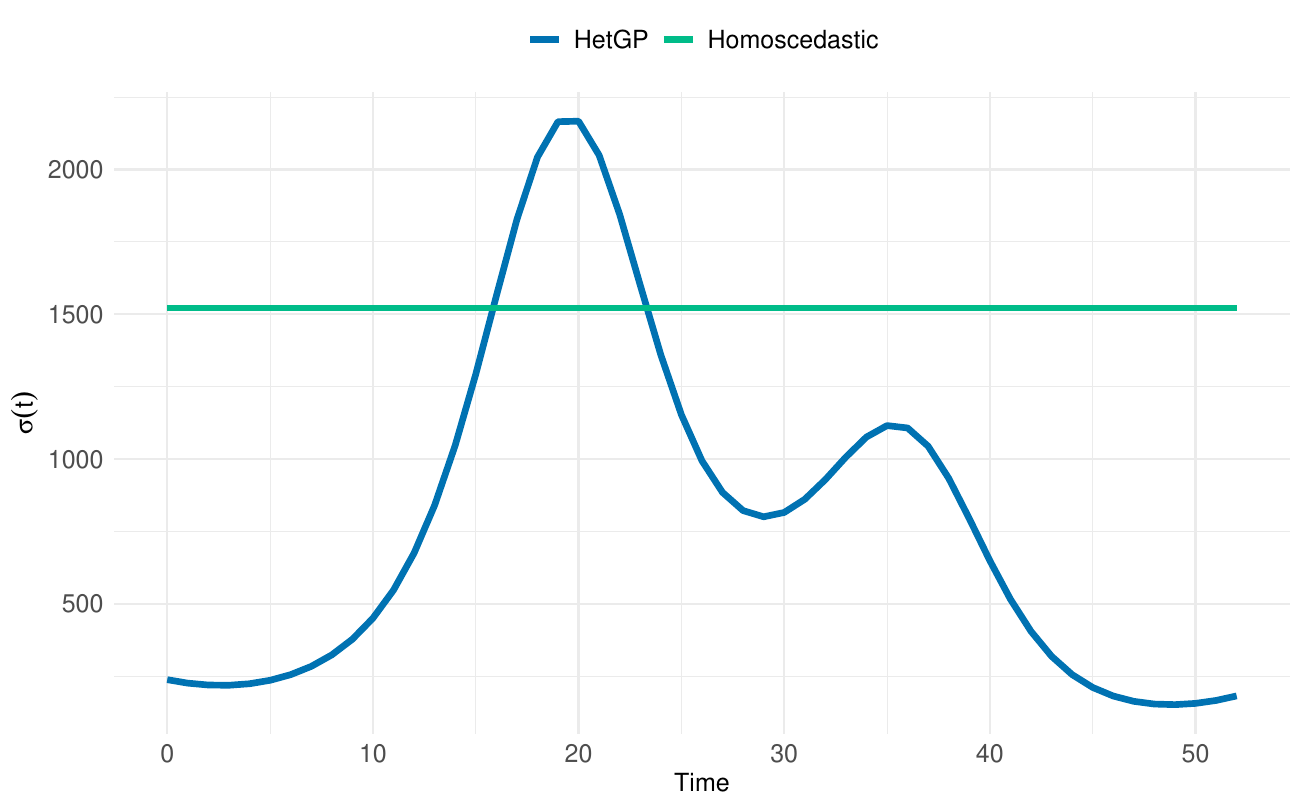}
\captionsetup{
    width=0.9\textwidth,
    font=footnotesize,  
    labelfont=bf,      
    justification=justified,
    skip=6pt         
}
\caption{Estimated observation error for the measles epidemic data under the HetGP model (blue), and a homoscedastic model (green).}
\label{fig:measles_sigma}
\end{figure}

The estimated observation error for the measles data is shown in Figure~\ref{fig:measles_sigma}. The HetGP model captures substantial temporal variation in measurement variability, with higher variance during epidemic peaks and lower variance during troughs. As in the previous examples, the automatic model selection option in \texttt{mleHetGP} yielded results identical to the forced heteroscedastic fit, suggesting that the heteroscedastic noise model was preferred according to the marginal likelihood criterion.

\begin{figure}[ht!]
\centering
\includegraphics[width=1\textwidth]{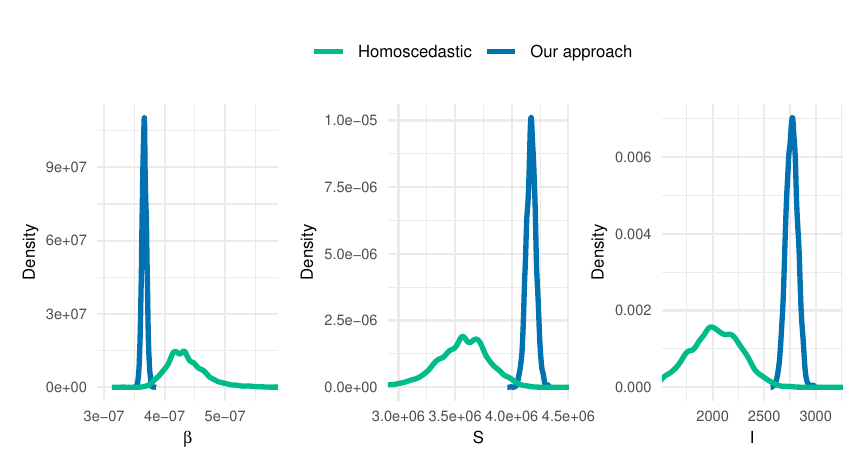}
\captionsetup{
    width=0.9\textwidth,
    font=footnotesize,  
    labelfont=bf,      
    justification=justified,
    skip=6pt         
}
\caption{Marginal posterior distributions for key epidemiological parameters. The blue curves correspond to our approach and the green curves
to the homoscedastic model. Our approach yields posterior estimates that differ from the homoscedastic case and are more consistent with the observed epidemic timing and scale under the HetGP-estimated error model.
}
\label{fig:measles_post}
\end{figure}

Figure~\ref{fig:measles_post} shows the resulting posterior distributions. There is a substantial difference between the marginal posterior estimates for our approach and the homoscedastic error assumption. 

\begin{figure}[ht!]
\centering
\includegraphics[width=1\textwidth]{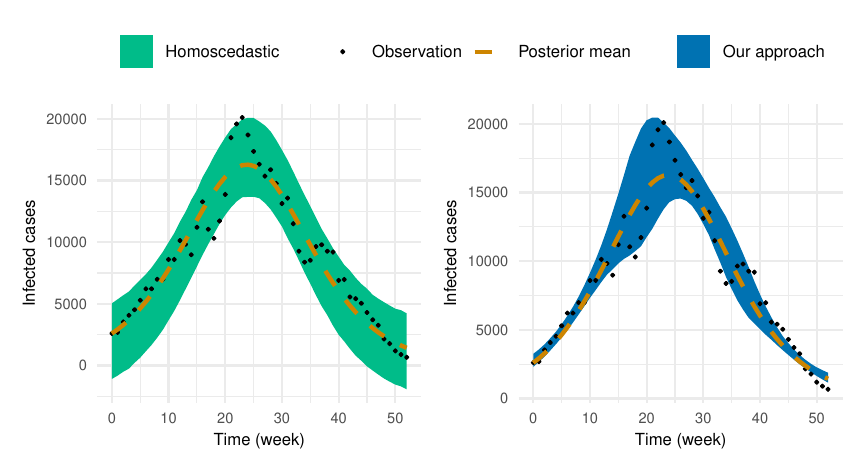}
\captionsetup{
    width=0.9\textwidth,
    font=footnotesize,  
    labelfont=bf,      
    justification=justified,
    skip=6pt         
}
\caption{Posterior predictive interval for weekly measles incidence (53 week subset). Observed cases are shown as points. Blue: our approach; green: homoscedastic model. Areas indicate $95\%$ credible intervals.}
\label{fig:measles_fit}
\end{figure}

While we cannot directly assess which posterior is closer to the truth, the posterior predictive intervals (Figure~\ref{fig:measles_fit}) under our approach provide a more accurate description of the data. While both capture the recurrent epidemic cycles, our approach produces credible intervals that expand near epidemic peaks and contract during troughs, consistent with the variance estimates from Step~1. 

This behaviour arises because infection counts exhibit lower variability when case numbers are small and higher variability near epidemic peaks. The heteroscedastic error model captures this pattern by assigning smaller observation variance at early time points and larger variance when infection counts increase. This leads to stronger constraints on the estimates of the initial susceptible population and the transmission rate, resulting in more concentrated posterior distributions for the epidemiological parameters. This in turn suggests that our framework yields a more reliable posterior approximation.

To provide an epidemiological interpretation of the inferred transmission parameters, we computed the posterior distribution of the basic reproduction number $R_0$. 
In epidemic modeling \citep{GUERRA2017e420}, $R_0$ is defined as the expected number of secondary infections generated by a typical infectious individual in a fully susceptible population, and serves as a threshold parameter determining whether sustained transmission can occur. When $R_0 > 1$, sustained transmission is possible, whereas $R_0 < 1$ implies not every case will result in a new infection in another individual, and outbreaks will eventually die out, although limited transmission chains may still occur.

\begin{figure}[ht!]
\centering
\includegraphics[width=1\textwidth]{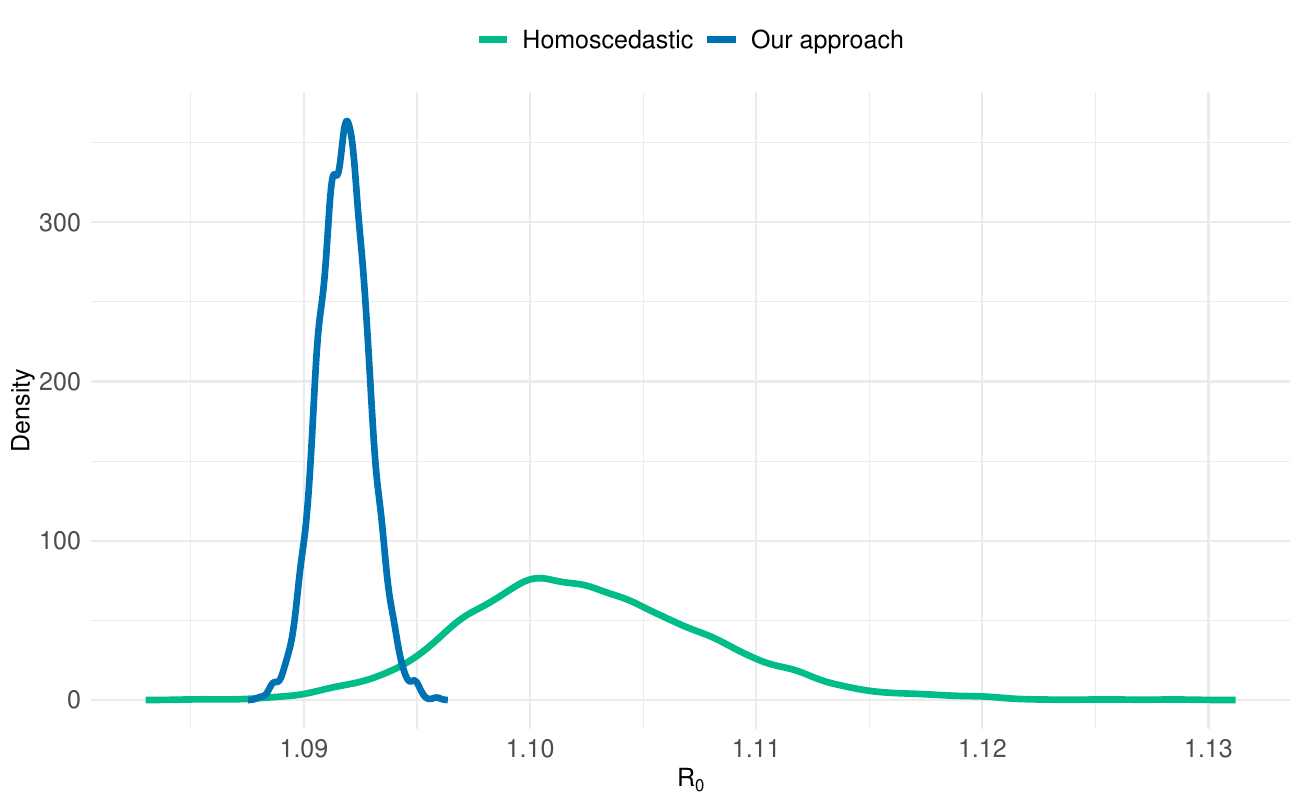}
\captionsetup{
    width=0.9\textwidth,
    font=footnotesize,  
    labelfont=bf,      
    justification=justified,
    skip=6pt         
}
\caption{Posterior distributions of the basic reproduction number $R_0$ under different error assumptions. The blue curve correspond to our approach and the green curve to the homoscedastic model.
}
\label{fig:R0}
\end{figure}

For the SIR system in Eq.~\eqref{Eq.measles}, the basic reproduction number is given by
\[
R_0 = \frac{\beta ( S_0+I_0)}{\delta},
\]
with $\delta = 7/5$. Posterior samples of $\beta$, $S_0$ and $I_0$ were substituted into this expression to obtain the induced posterior distribution of $R_0$ under both modeling assumptions. Figure~\ref{fig:R0} shows the posterior distributions of $R_0$ under the two modeling assumptions. 
The homoscedastic model yields a broader posterior centred around slightly larger values of $R_0$, 
whereas our approach produces a more concentrated posterior distribution at lower values. Although previous studies using this dataset do not report $R_0$ explicitly, the parameter estimates obtained here are broadly consistent with those reported in earlier analyses of the same data.

\FloatBarrier

\section{Discussion}
\label{sec:discussion}
This work demonstrates that explicitly modeling heteroscedastic measurement error can substantially influence parameter inference in ODE models. Across all case studies, allowing the observation variance to vary over time produced predictive intervals that were more consistent with the variability evident in the data than those obtained under a constant-variance assumption. In the logistic simulation, where the true variance structure is known, the heteroscedastic approach yielded posterior samples closer to the benchmark posterior as the number of observations increased and produced predictive intervals that adapt to the time-varying error level. In the coral and measles examples, the HetGP stage identified pronounced changes in observational variability over time, and incorporating these estimates into the likelihood led to meaningful changes in marginal posterior distributions and to predictive intervals that expand and contract in line with the inferred error process. These results highlight that the incorrect assumptions about the error structure, such as assuming homoscedasticity, can lead to misleading inferences and yield overconfident, or biased parameter estimation.

A key feature of the proposed framework is the separation of stochastic error modeling from mechanistic parameter inference. In Step~1, a HetGP is used to learn a flexible, time-dependent error variance without requiring a parametric variance specification. In Step~2, these variance estimates are incorporated into a heteroscedastic likelihood for the ODE model parameters (Algorithm~\ref{alg:Method}), yielding a practical workflow that avoids the computational and identifiability challenges associated with simultaneously learning both the ODE dynamics and a highly flexible error process. From an applied perspective, this separation makes the approach easy to deploy, the error structure can be diagnosed and learned directly from the data, and the resulting variance function can be used as an informed component of the observation model in Bayesian inference.

Nonetheless, a key limitation of the present approach is its reliance on Gaussian assumptions for the measurement error. While this choice is convenient and analytically tractable, empirical data may exhibit characteristics that depart substantially from Gaussianity, including skewness, heavy tails, or small discrete counts. While the proposed framework is general, our current implementation is limited to Gaussian error models with heteroscedastic variance. Future work could extend the implementation to other types of error models or alternative error structures, which are often encountered in practice \citep{Lambert2023,NANSHAN2022107483,Warne_2025}. Relaxing the Gaussianity assumption could be achieved by adopting flexible non-Gaussian error models, such as generalized additive models (GAMs) \citep{10.1214/ss/1177013604}, which retain interpretability while accommodating diverse observation distributions. 

Overall, our findings support the view that careful modeling of the measurement process is as critical as specifying the dynamical model itself. Incorporating heteroscedasticity provides more reliable uncertainty quantification. Future research should explore extensions to non-Gaussian error to improve inference robustness, thereby moving towards a more comprehensive framework for Bayesian inference in ODE systems.

\section*{Statements and Declarations}
\textbf{Funding}

SS acknowledges support from the Queensland University of Technology through the QUT Postgraduate Research Award (QUTPRA).
CD was supported by an Australian Research Council (ARC) Future Fellowship (FT210100260). 
DJW was supported by an ARC Discovery Early Career Researcher Award (DE250100396). 
CD and DJW also acknowledge support from the ARC Centre of Excellence for the Mathematical Analysis of Cellular Systems (MACSYS) (CE230100001).

\textbf{Data and Code availability}

 All code and data required to reproduce the results in this work are publicly available at
\href{https://github.com/SelvaSalimi/input-dependent-noise}{GitHub repository}.

\textbf{Competing Interests}

The authors declare no competing interests.
\newpage

\bibliographystyle{abbrvnat}
\bibliography{CRef}

@Article{Alahmadi2020,
  author    = {Alahmadi, Amani A. and Flegg, Jennifer A. and Cochrane, Davis G. and Drovandi, Christopher C. and Keith, Jonathan M.},
  journal   = {Royal Society Open Science},
  title     = {A comparison of approximate versus exact techniques for {B}ayesian parameter inference in nonlinear ordinary differential equation models},
  year      = {2020},
  issn      = {2054-5703},
  number    = {3},
  pages     = {191315},
  volume    = {7},
  doi       = {10.1098/rsos.191315},
  publisher = {The Royal Society},
}

@article{Binois_2018,
   title={Practical Heteroscedastic {G}aussian Process Modeling for Large Simulation Experiments},
   volume={27},
   ISSN={1537-2715},
   DOI={10.1080/10618600.2018.1458625},
   number={4},
   journal={Journal of Computational and Graphical Statistics},
   publisher={Informa UK Limited},
   author={Binois, Mickaël and Gramacy, Robert B. and Ludkovski, Mike},
   year={2018}, pages={808–821} 
}

@article{article_HetGP,
author = {Binois, Mickaël and Gramacy, Robert},
year = {2021},
pages = {},
title = {{H}et{GP}: Heteroskedastic {G}aussian Process Modeling and Sequential Design in {R}},
volume = {98},
journal = {Journal of Statistical Software},
doi = {10.18637/jss.v098.i13}
}

@Article{Browning2020,
  author    = {Browning, Alexander P. and Warne, David J. and Burrage, Kevin and Baker, Ruth E. and Simpson, Matthew J.},
  journal   = {Journal of The Royal Society Interface},
  title     = {Identifiability analysis for stochastic differential equation models in systems biology},
  year      = {2020},
  issn      = {1742-5662},
  number    = {173},
  pages     = {20200652},
  volume    = {17},
  doi       = {10.1098/rsif.2020.0652},
  publisher = {The Royal Society},
}

@article{Dattner2015AMI,
  title={A model‐based initial guess for estimating parameters in systems of ordinary differential equations},
  author={Itai Dattner},
  journal={Biometrics},
  year={2015},
  volume={71}
}

@article{DAVIDZON20125397,
title = {Newton’s law of cooling and its interpretation},
journal = {International Journal of Heat and Mass Transfer},
volume = {55},
number = {21},
pages = {5397-5402},
year = {2012},
issn = {0017-9310},
doi = {10.1016/j.ijheatmasstransfer.2012.03.035},
author = {Michael I. Davidzon}
}

@Book{Gelman_2013,
  author    = {Gelman, Andrew and Carlin, John B. and Stern, Hal S. and Dunson, David B. and Vehtari, Aki and Rubin, Donald B.},
  publisher = {Chapman and Hall/CRC},
  title     = {{B}ayesian Data Analysis},
  year      = {2013},
  isbn      = {9780429113079},
  doi       = {10.1201/b16018},
}

@inproceedings{NIPS1997_afe43465,
 author = {Goldberg, Paul and Williams, Christopher and Bishop, Christopher},
 booktitle = {Advances in Neural Information Processing Systems},
 editor = {M. Jordan and M. Kearns and S. Solla},
 pages = {493-499},
 publisher = {MIT Press},
 title = {Regression with Input-dependent Noise: A {G}aussian Process Treatment},
 volume = {10},
 year = {1997}
}

@article{JMLR:v13:gretton12a,
  author  = {Arthur Gretton and Karsten M. Borgwardt and Malte J. Rasch and Bernhard Sch{{\"o}}lkopf and Alexander Smola},
  title   = {A Kernel Two-Sample Test},
  journal = {Journal of Machine Learning Research},
  year    = {2012},
  volume  = {13},
  number  = {25},
  pages   = {723-773}
}

@article{GUERRA2017e420,
title = {The basic reproduction number ({R}0) of measles: a systematic review},
journal = {The Lancet Infectious Diseases},
volume = {17},
number = {12},
pages = {e420-e428},
year = {2017},
issn = {1473-3099},
doi = {10.1016/S1473-3099(17)30307-9},
author = {Fiona M Guerra and Shelly Bolotin and Gillian Lim and Jane Heffernan and Shelley L Deeks and Ye Li and Natasha S Crowcroft}
}

@article{10.1214/ss/1177013604,
author = {Trevor Hastie and Robert Tibshirani},
title = {{Generalized Additive Models}},
volume = {1},
journal = {Statistical Science},
number = {3},
publisher = {Institute of Mathematical Statistics},
pages = {297-310},
year = {1986},
doi = {10.1214/ss/1177013604}
}

@article{2020_Huang,
author = {Hanwen Huang and Andreas Handel and Xiao Song},
title = {A {B}ayesian approach to estimate parameters of ordinary differential equation},
journal = {Computational Statistics},
year = {2020},
volume = {35},
publisher = {Springer Nature},
number = {3},
pages = {1481--1499},
doi = {10.1007/s00180-020-00962-8}
}

@Article{RePEc:bla:jorssc:v:71:y:2022:i:5:p:1978-1995,
  author={Hanwen Huang},
  title={{B}ayesian multi‐level mixed‐effects model for influenza dynamics},
  journal={Journal of the Royal Statistical Society Series C},
  year=2022,
  volume={71},
  number={5},
  pages={1978-1995},
  doi={10.1111/rssc.12603}
}

@Article{Lambert2023,
  author    = {Lambert, Ben and Lei, Chon Lok and Robinson, Martin and Clerx, Michael and Creswell, Richard and Ghosh, Sanmitra and Tavener, Simon and Gavaghan, David J.},
  journal   = {Journal of The Royal Society Interface},
  title     = {Autocorrelated measurement processes and inference for ordinary differential equation models of biological systems},
  year      = {2023},
  issn      = {1742-5662},
  number    = {199},
  volume    = {20},
  doi       = {10.1098/rsif.2022.0725},
  publisher = {The Royal Society},
}

@Article{Maclaren2017,
  author    = {Oliver J. Maclaren and Aim{\'{e}}e Parker and Carmen Pin and Simon R. Carding and Alastair J. M. Watson and Alexander G. Fletcher and Helen M. Byrne and Philip K. Maini},
  journal   = {PLOS Computational Biology},
  title     = {A hierarchical {B}ayesian model for understanding the spatiotemporal dynamics of the intestinal epithelium},
  year      = {2017},
  number    = {7},
  volume    = {13},
  bibsource = {dblp computer science bibliography, https://dblp.org},
  doi       = {10.1371/JOURNAL.PCBI.1005688},
}

@article{10.1063/1.1699114,
    author = {Metropolis, Nicholas and Rosenbluth, Arianna W. and Rosenbluth, Marshall N. and Teller, Augusta H. and Teller, Edward},
    title = {Equation of State Calculations by Fast Computing Machines},
    journal = {The Journal of Chemical Physics},
    volume = {21},
    number = {6},
    pages = {1087-1092},
    year = {1953},
    issn = {0021-9606},
    doi = {10.1063/1.1699114}
}

@Article{Murphy2022,
  author    = {Murphy, Ryan J. and Maclaren, Oliver J. and Calabrese, Alivia R. and Thomas, Patrick B. and Warne, David J. and Williams, Elizabeth D. and Simpson, Matthew J.},
  journal   = {Journal of The Royal Society Interface},
  title     = {Computationally efficient framework for diagnosing, understanding and predicting biphasic population growth},
  year      = {2022},
  issn      = {1742-5662},
  number    = {197},
  volume    = {19},
  doi       = {10.1098/rsif.2022.0560},
  publisher = {The Royal Society},
}

@article{NANSHAN2022107483,
title = {Dynamical modeling for non-{G}aussian data with high-dimensional sparse ordinary differential equations},
journal = {Computational Statistics \& Data Analysis},
volume = {173},
pages = {107483},
year = {2022},
issn = {0167-9473},
doi = {10.1016/j.csda.2022.107483},
author = {Muye Nanshan and Nan Zhang and Xiaolei Xun and Jiguo Cao}
}

@article{PILLONETTO2014657,
title = {Kernel methods in system identification, machine learning and function estimation: A survey},
journal = {Automatica},
volume = {50},
number = {3},
pages = {657-682},
year = {2014},
issn = {0005-1098},
doi = {10.1016/j.automatica.2014.01.001},
author = {Gianluigi Pillonetto and Francesco Dinuzzo and Tianshi Chen and Giuseppe {De Nicolao} and Lennart Ljung}
}

@book{Rasmussen2006Gaussian,
  author = {Rasmussen, Carl Edward and Williams, Christopher K. I.},
  publisher = {The MIT Press},
  title = {{G}aussian Processes for Machine Learning},
  year = 2006
}

@article{10.1093/jxb/10.2.290,
    author = {Richards, F. J.},
    title = {A Flexible Growth Function for Empirical Use},
    journal = {Journal of Experimental Botany},
    volume = {10},
    number = {2},
    pages = {290-301},
    year = {1959},
    month = {06},
    issn = {0022-0957},
    doi = {10.1093/jxb/10.2.290}
}

@article{Robert_2011,
   title={A Short History of {M}arkov Chain {M}onte {C}arlo: Subjective Recollections from Incomplete Data},
   volume={26},
   ISSN={0883-4237},
   DOI={10.1214/10-sts351},
   number={1},
   journal={Statistical Science},
   publisher={Institute of Mathematical Statistics},
   author={Robert, Christian and Casella, George},
   year={2011} 
}

@article{c56b4e38-c4b3-34ac-8e24-87699314559b,
 ISSN = {10505164},
 author = {G. O. Roberts and A. Gelman and W. R. Gilks},
 journal = {The Annals of Applied Probability},
 number = {1},
 pages = {110--120},
 publisher = {Institute of Mathematical Statistics},
 title = {Weak Convergence and Optimal Scaling of Random Walk {M}etropolis Algorithms},
 urldate = {2025-07-20},
 volume = {7},
 year = {1997}
}

@article{SIMPSON2022110998,
title = {Parameter identifiability and model selection for sigmoid population growth models},
journal = {Journal of Theoretical Biology},
volume = {535},
pages = {110998},
year = {2022},
issn = {0022-5193},
doi = {10.1016/j.jtbi.2021.110998},
author = {Matthew J. Simpson and Alexander P. Browning and David J. Warne and Oliver J. Maclaren and Ruth E. Baker}
}

@article{TSOULARIS200221,
title = {Analysis of logistic growth models},
journal = {Mathematical Biosciences},
volume = {179},
number = {1},
pages = {21-55},
year = {2002},
issn = {0025-5564},
doi = {10.1016/S0025-5564(02)00096-2},
author = {A. Tsoularis and J. Wallace}
}

@article{https://doi.org/10.1111/1365-2664.14039,
author = {Warne, David J. and Crossman, Kerryn A. and Jin, Wang and Mengersen, Kerrie and Osborne, Kate and Simpson, Matthew J. and Thompson, Angus A. and Wu, Paul and Ortiz, Juan-C.},
title = {Identification of two-phase recovery for interpretation of coral reef monitoring data},
journal = {Journal of Applied Ecology},
volume = {59},
number = {1},
pages = {153-164},
doi = {10.1111/1365-2664.14039},
year = {2022}
}

@article{Warne_2025,
   title={Mathematical Modelling and Uncertainty Quantification for Analysis of Biphasic Coral Reef Recovery Patterns},
   volume={87},
   ISSN={1522-9602},
   DOI={10.1007/s11538-025-01512-3},
   number={9},
   journal={Bulletin of Mathematical Biology},
   publisher={Springer Science and Business Media LLC},
   author={Warne, David J. and Crossman, Kerryn and Heron, Grace E. M. and Sharp, Jesse A. and Jin, Wang and Wu, Paul Pao-Yen and Simpson, Matthew J. and Mengersen, Kerrie and Ortiz, Juan-C.},
   year={2025} 
}

@article {Warne2025.12.21.695338,
	author = {Warne, David J. and Zhu, Xiangrun and Steele, Thomas P. and Johnston, Stuart T. and Sisson, Scott A. and Faria, Matthew and Murphy, Ryan J. and Browning, Alexander P.},
	title = {A multilevel hierarchical framework for quantification of experimental heterogeneity},
	elocation-id = {2025.12.21.695338},
	year = {2025},
	doi = {10.64898/2025.12.21.695338},
	publisher = {Cold Spring Harbor Laboratory},
	journal = {bioRxiv}
}

@inproceedings{NIPS1995_7cce53cf,
 author = {Williams, Christopher and Rasmussen, Carl},
 booktitle = {Advances in Neural Information Processing Systems},
 editor = {D. Touretzky and M.C. Mozer and M. Hasselmo},
 pages = {514-520},
 publisher = {MIT Press},
 title = {Gaussian Processes for Regression},
 volume = {8},
 year = {1995}
}

@article{XIA2026117428,
title = {Some application of ordinary differential equations in several areas of the economy},
journal = {Chaos, Solitons \& Fractals},
volume = {202},
pages = {117428},
year = {2026},
issn = {0960-0779},
doi = {10.1016/j.chaos.2025.117428},
author = {Hongmei Xia and Meirong Yang and Guoxiang Han and Yuchuan Huang and Yudong Fang and Dongyue Meng}
}

@Inbook{Potter2023,
author="Potter, Merle C.
and Feeny, Brian F.",
title="Ordinary Differential Equations",
bookTitle="Mathematical Methods for Engineering and Science",
year="2023",
publisher="Springer International Publishing",
address="Cham",
pages="1--75",
isbn="978-3-031-26151-0",
doi="10.1007/978-3-031-26151-0_1"
}

@article{hetarticle,
author = {Ohaegbulem, Emmanuel and Iheaka, Victor},
year = {2024},
pages = {225-261},
title = {On Remedying the Presence of Heteroscedasticity in a Multiple Linear Regression Modelling},
volume = {7},
journal = {African Journal of Mathematics and Statistics Studies},
doi = {10.52589/AJMSS-TJ9XI8HD}
}
\end{document}